\documentclass[twocolumn,preprintnumbers,amssymb,amsmath,superscriptaddress,letterpaper,nofootinbib]{revtex4}[12pt]
\usepackage{times,graphics}
\usepackage[colorlinks]{hyperref}
\usepackage{lmodern}
\usepackage{graphicx}
\usepackage{dcolumn}
\usepackage{bm}
\usepackage{pstricks}
\usepackage{epsfig}
\usepackage{epstopdf}
\usepackage{multirow}
\usepackage{amsmath}
\usepackage{lipsum}
\usepackage{lineno}
\begin{document}
\title{A Semiblind Reconstruction of the History of Effective Number of Neutrinos Using CMB Data}

\author{Sarah Safi}
\affiliation{Department of Physics, Shahid Beheshti University, 1983969411,  Tehran Iran}
\author{Marzieh Farhang}
\affiliation{Department of Physics, Shahid Beheshti University, 1983969411,  Tehran Iran}

\author{Olga Mena}
\affiliation{Instituto de F{\'\i}sica Corpuscular  (CSIC-Universitat de Val{\`e}ncia), E-46980 Paterna, Spain}

\author{Eleonora Di Valentino}  
\affiliation{School of Mathematics and Statistics, University of Sheffield, Hounsfield Road, Sheffield S3 7RH, United Kingdom}

\date{\today}% It is always \today, today,
             %  but any date may be explicitly specified

\begin{abstract}
We explore the possibility of redshift-dependent deviations in the contribution of relativistic degrees of freedom to the radiation budget of the cosmos, conventionally parameterized by the effective number of neutrinos $N_{\rm eff}$, from the predictions of the standard model. 
We expand the deviations $\Delta N_{\rm eff}(z)$ in terms of top-hat functions and treat their amplitudes as the free parameters of the theory to be measured alongside the standard cosmological parameters by the \textit{Planck} measurements of the cosmic microwave background (CMB) anisotropies and Baryonic Acoustic Oscillations, as well as performing forecasts for futuristic CMB surveys such as PICO and CMB-S4. 
We reconstruct the history of $\Delta N_{\rm eff}$ and find that with the current data the history is consistent with the standard scenario. 
Inclusion of the new degrees of freedom in the analysis increases $H_0$ to $68.71\pm 0.44$, slightly reducing the Hubble tension. With the smaller forecasted errors on the $\Delta N_{\rm eff}(z)$ parametrization modes from future CMB surveys, very accurate bounds are expected within the possible range of dark radiation models.
\end{abstract}


\maketitle

\section{Introduction}\label{sec:background}
In the standard model of cosmology, the contribution of cosmic neutrinos to the radiation budget of the Universe (in epochs where their temperature is well above their mass) is commonly parameterized by the effective number of neutrinos, $N_{\rm eff}$, through
\begin{equation}\label{eq:neff}
\rho _{\rm r}=\rho _{\gamma} \Big [1+\frac{7}{8}(\frac{4}{11})^{4/3}N_{\rm eff}\Big ]~,
\end{equation}
where $\rho _{\rm r}$ is the total energy density in relativistic species, $\rho _{\gamma}$ is the photon energy density in the form of the Cosmic Microwave Background radiation (CMB), and $N_{\rm eff}=3.044$ corresponds to the prediction within the standard model~\citep{Akita:2020szl,Froustey:2020mcq,Bennett:2020zkv}. 
A value of $\Delta N_{\rm{eff}} ( \equiv  N_{\rm{eff}} -3.044) >0$ would thus imply extra contributions from either non-interacting or weakly interacting relativistic species, often referred to as dark radiation \citep[see, e.g., ][]{AbazajianCarlstromLee:2013, calabrese2011limits,Archidiacono:2011gq}.
Given the weak (if any) coupling of  these candidates to other particles of the standard model, these particles are hard to produce in laboratories or to be directly detected in Earth-bound detectors. 
Cosmological observations therefore play a unique role in exploring the various extensions to the standard model of particle physics through the gravitational impact of these dark degrees of freedom on the matter distribution and CMB anisotropies.
Among the different models proposed are sterile neutrinos \citep{Joudaki:2012uk,Jacques:2013xr,Gariazzo:2015rra,Archidiacono:2022ich}, gravitons produced from Hawking evaporation of spinning primordial black holes in the early Universe \citep{Arbey:2021ysg},  Dirac neutrinos with thermalized right-handed components \citep{Abazajian:2019oqj,Luo:2020sho} and thermal axions \citep{Conlon:2013isa,Baumann:2016wac}.
%-----------------------------------------------------
\begin{figure*}
\centering
 \includegraphics[width=\columnwidth]{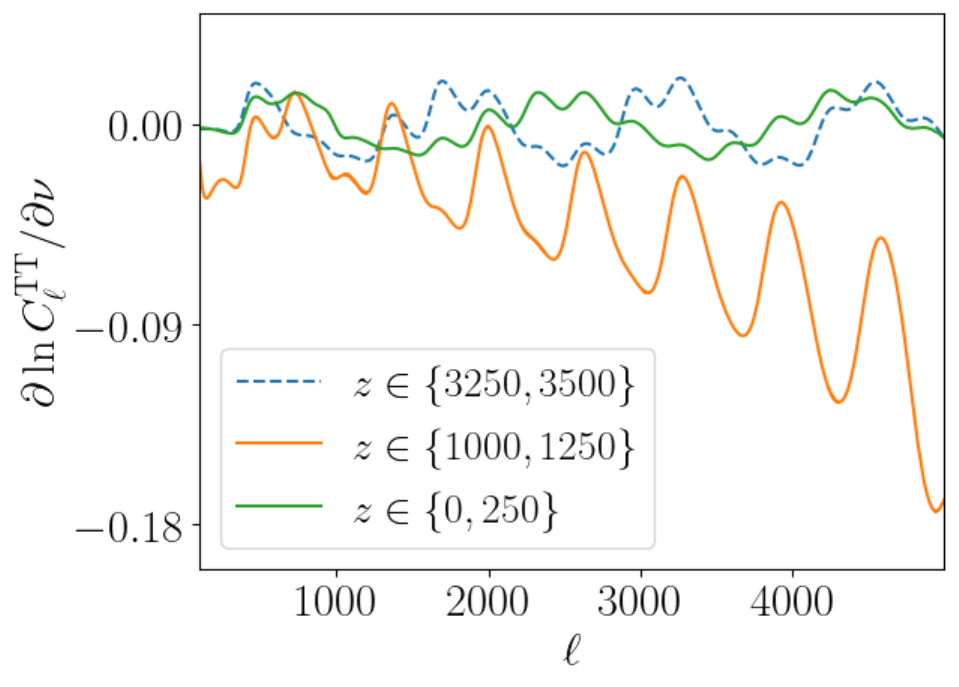}
  \includegraphics[width=\columnwidth]{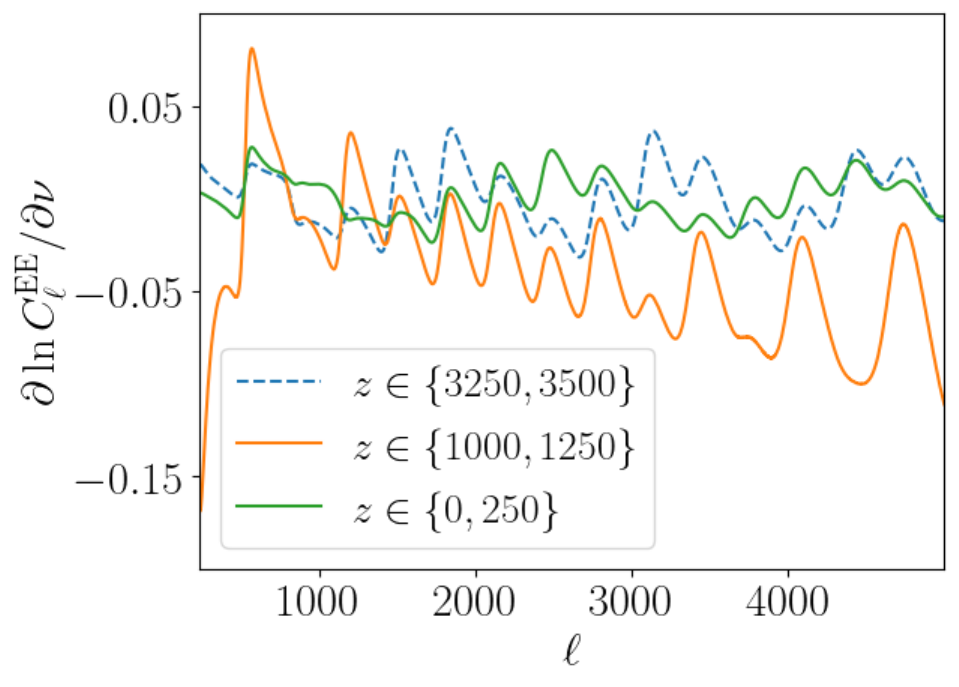}
 \caption{The sensitivity of $C^{\rm TT}_{\ell}$ (left) and $C^{\rm EE}_{\ell}$(right) to $\Delta N_{\rm{eff}}$ in the form of top-hats for three different redshift bins.}
 \label{fig:bin_sensitivity}
\end{figure*}
A smaller value of $N_{\rm eff}$, on the other hand, is harder to interpret and may require modifications to the thermal history, such as energy injection to the CMB after neutrino decoupling or reheating at very low tempratures \citep{Gelmini:2004ah,finkbeiner2012searching,Poulin:2016anj,Freese:2017ace}. The extra energy density in these free-streaming dark degrees of freedom would affect the expansion rate at and prior to recombination as well as redshift of matter-radiation equality, $z_{\rm eq}$. The increased $H(z)$ would in turn impact the angular size of the sound horizon at recombination, $\theta_{\rm s}=r_{\rm s}/D_{\rm A}$, where $r_{\rm s}=\int ^{a_*}_{0}{c_{\rm s}{\rm d}a}/{a^2H}$, $D_{\rm A}=\int ^1_{a_*} c{\rm d}t/a$ and  ${a_*}$ denotes the scale factor at recombination. As these quantities (i.e., $\theta_{\rm s}$ and $z_{\rm eq}$) are well measured by data, changes in $N_{\rm eff}$ should be compensated by appropriate variations in other cosmological parameters, such as those characterizing late time geometry and matter density. Yet, even if the impact of $\Delta N_{\rm eff}$ on the power spectrum at low $\ell$ is minimized through holding $z_{\rm eq}$ and $\theta_{\rm s}$ fixed, higher $N_{\rm eff}$ would lead to enhanced damping of CMB power spectrum at scales smaller than the photon diffusion length. This makes the high-$\ell$ CMB power spectrum sensitive to variations in $N_{\rm eff}$.

CMB measurements from \cite{pl18} have led to $N_{\rm eff}=2.92\pm 0.36$ at 95\% confidence level (CL), consistent with theoretical predictions (with an insignificant bias toward lower values). More accurate measurements of $N_{\rm eff}$ with future experiments such as PICO \citep{pico} and CMB-S4 \citep{CMB-S4} are crucial to strongly confirm the standard model prediction or to open paths towards new physics.

In this work our goal is to explore possible redshift-dependent variations in $N_{\rm eff}$ and provide a model-independent reconstruction of the history of this parameter \citep[see][for a model independent reconstruction of neutrino mass]{Lorenz:2021alz}. 
We allow the contribution to the total radiation density of the Universe from the non-CMB component to deviate from its standard model value and search for the most constrainable patterns in this deviation history through a semi-blind approach.
A thorough analysis would require coupling of these relativistic species to other energy sectors in the cosmos to satisfy energy conservation. 
Here we  assume those possible couplings  have negligible  impact on the cosmological observables.

The paper is organized as follows. 
In Section~\ref{sec:methodology} we explain the details of the methodology used for the redshift-dependent parametrization of $N_{\rm eff}$, the process of eigenmode reconstruction for the $N_{\rm eff}$ history, and the data and simulations used for the analysis. The results of the analysis applied to current CMB data and the predictions based on futuristic  CMB simulations are presented in Section \ref{sec:results}. We conclude in Section \ref{sec:conclusion}.
\section{Methodology}\label{sec:methodology}
In this section, we explain the method used  to model and  investigate redshift-dependent deviations in the effective number of neutrinos around the predictions of the standard model.
%-----------------------------------------------------
\subsection{$\Delta N_{\rm {eff}}(z)$ Parametrization}\label{parametrization}
As discussed in the introduction, the effective number of relativistic degrees of freedom in the Universe is characterized by the single parameter $N_{\rm{eff}}$.
To search for extensions to the standard model with an impact on this degree of freedom, we allow for redshift-dependent deviations of $N_{\rm{eff}}$ around its theoretical prediction of $3.044$.
Specifically, we expand the parameter space describing the relativistic species by treating $N_{\rm{eff}}$ as a redshift-dependent parameter binned into $n_{\rm b}$ linearly spaced redshift bins for $z \in \{ z_{\rm min}, z_{\rm max} \}$. 
The bin amplitudes are the new  parameters which we label by $\{\nu_1,\cdots,\nu_{n_{\rm b}}\} $. The $N_{\rm{eff}}$ history can therefore be approximated by $\Delta N_{\rm eff}(z)\approx \sum _{i=1}^{n_{\rm b}} \nu _i T_i(z)$, where $T_i(z)$ represents the top-hat function corresponding to the $i$th bin. The $T_i(z)$'s  form an orthonormal basis in the redshift range $z \in \{ z_{\rm min}, z_{\rm max} \}$, if the bin widths are small compared to the shortest redshift scales explored in the problem.
The goal is to measure the $\nu_i$'s, alongside the parameters of the standard model $\{\Omega _bh^2,\Omega _ch^2,\tau,H_0,n_s,\ln (10^{10} A_s)\}$, using cosmological data. 
 To achieve this, we modified the publicly available code CosmoMC\footnote{\url{https://cosmologist.info/cosmomc/}}
\citep{lewis02,lewis13} to include the $\Delta N_{\rm{eff}}$ parameters and utilize various combinations of current and futuristic datasets to search for possible deviations from the standard model. 
%----------------------------------------------
%-----------------------------------------------------------------
\subsection{$\Delta N_{\rm eff}(z)$ Eigenmodes}
The measured amplitudes of the  $\Delta N_{\rm eff}$ bins can in principle be used to construct the  $N_{\rm eff}$ history. However, one might expect relatively large errors in their measurements due to possible substantial correlations between redshift bins.
This could be the case in particular for some neighboring bins whose impacts on data are hardly distinguishable from one to another.  
The large errors would, in turn, lead to poor reconstruction of the $N_{\rm eff}$ history and hide the potentially invaluable information hidden in the data about the number of relativistic species in the cosmos. 
Figure \ref{fig:bin_sensitivity} illustrates the sensitivity of the CMB temperature power spectrum to $\Delta N_{\rm eff}(z)$ for different redshift bins. We notice that the largest derivative amplitude happens for $z\in \{1000,1250\}$, i.e., around the epoch of recombination, particularly at high multipoles.
 \begin{figure*}
   \begin{center}
\makebox[\textwidth][c]{\includegraphics[width=\textwidth]{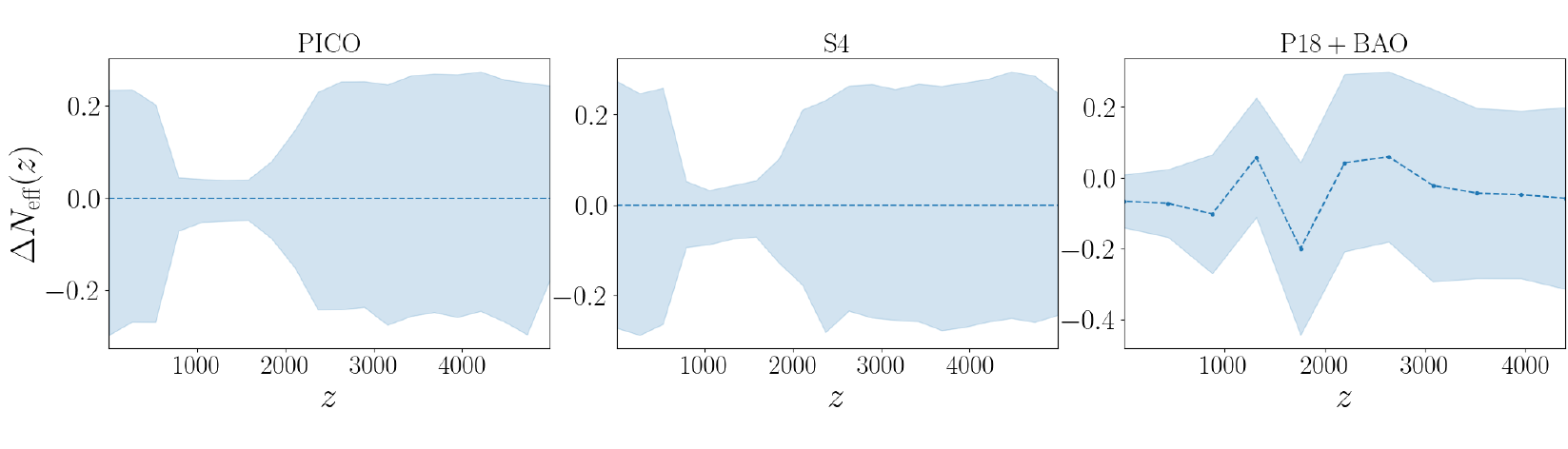}}%
    \caption{The reconstruction of $N_{\rm eff}$ history with top-hats as  basis functions. The blue dashed lines are the mean and the shaded areas represent the $1\sigma$ uncertainty region around the mean. Left and middle: forecasts for PICO and CMB-S4, respectively, with 20 linearly-spaced redshift bins in $\{0,5000\}$.  Right: P18+BAO measurements with 11 redshift bins, linearly spaced in $\{0,4400\}$.  }
\label{fig:pico-s4-p18-bao}
\end{center}
\end{figure*}
A way out of this hurdle is to construct a set of  $n_{\rm b}$ linearly uncorrelated parameters as linear combinations of the initial $n_{\rm b}$ bin amplitudes and use them as the new parameters to be included in the analysis. 
In other words, from these parameter combinations, one builds a new set of orthonormal basis functions that can be used to represent $\Delta N_{\rm eff}(z)$ in a more appropriate way.
This data-driven parameter construction, also known as Principal Component Analysis (PCA), has been widely used  in different contexts in cosmology (see, e.g., \cite{Farhang_2012,Farhang_2013}  for recombination history, \cite{Pandolfi_2010,Villanueva_Domingo_2018} for reionization and \cite{Farhang_2019, Esmaeilian_2021} for primordial anisotropies). 

The principal modes that characterize the $N_{\rm eff}$ history are constructed from the eigenvectors of the covariance matrix of the $\nu _i$'s,$
{\bm C}=\langle \Delta {\bm \nu}\Delta {\bm \nu} \rangle
$
where $\Delta {\bm \nu }={\bm \nu } -\langle {\bm \nu } \rangle$,  $\bm \nu =(\nu _1,\cdots ,\nu _{n _{\rm b}})$ and the angular brackets denote the expectation value.
We label these eigenmodes by $\tilde{T}_i(z)$, $i=1,\cdots ,n_{\rm b}$, and their amplitudes, which are to be constrained, by $\tilde{\nu}_i$. The eigenfuctions $\tilde{T}_i(z)$ are, by construction, orthogonal to each other. They can therefore form an orthonormal basis to expand functions with relatively slow redshift variations compared to the shortest redshift scales probed by the eigenmodes (in turn, inherited from the bins). We thus have
\begin{equation}
\Delta N_{\rm eff}(z)\approx \sum _{i=1}^{n_{\rm b}}{\nu}_i {T}_i(z)\approx \sum _{i=1}^{n_{\rm b}}\tilde{\nu}_i \tilde{T}_i(z)~.
\end{equation}
The uncertainty in the measurement of $\tilde{\nu}_i$'s  is estimated from the square root of the eigenvalues of $\bold{C}$.

To construct the covariance matrix of ${\nu}_i$'s, we use the CosmocMC package, modified to include changes required by the extended $N_{\rm eff}(z)$ model. For analysis with P18 data, we use the \textit{Planck} likelihood package \citep{Planck:2019nip}, and for BAO data and S4 and PICO simulations we employ the CosmoMC built-in likelihood models. The resulting chains are then fed to GetDist for post-processing, and in particular to obtain an estimate of the parameter covariance matrix.
Note that we allow for  variations of the standard cosmological and nuisance parameters in the analysis, and therefore the uncertainty in their measurement is taken into account in the $N_{\rm eff}(z)$ eigenmode construction.

 A main advantage of using the new parameters in the analysis is that  the parameter hierarchy can be truncated based on their estimated errors.
In other words, one can keep the parameters with relatively low uncertainty and drop the rest, without impacting the parameter estimation (in linear order) given that the new parameters are uncorrelated. 
 For more details on the method, see, e.g., \cite{Farhang_2019}.   
%--------------------------------------------------
%--------------------------------------------------
\begin{figure*}
   \begin{center}
   \makebox[\textwidth][c]{\includegraphics[width=\textwidth]{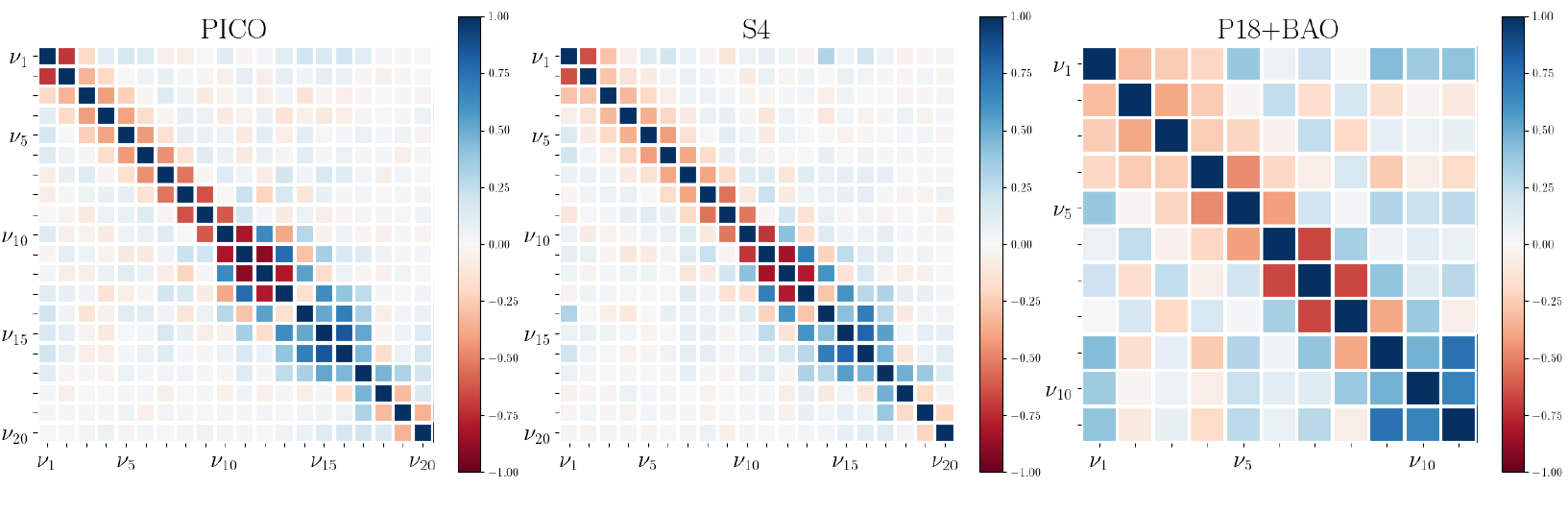}}%
    \caption{Correlation matrices for $\nu _i$s, for PICO and CMB-S4 (left and middle) and for P18+BAO   (right).}
    \label{fig:corrpico}
    \end{center}
\end{figure*}
\subsection{Data and Simulations}
\begin{figure}
    \centering
        \includegraphics[width=\columnwidth]{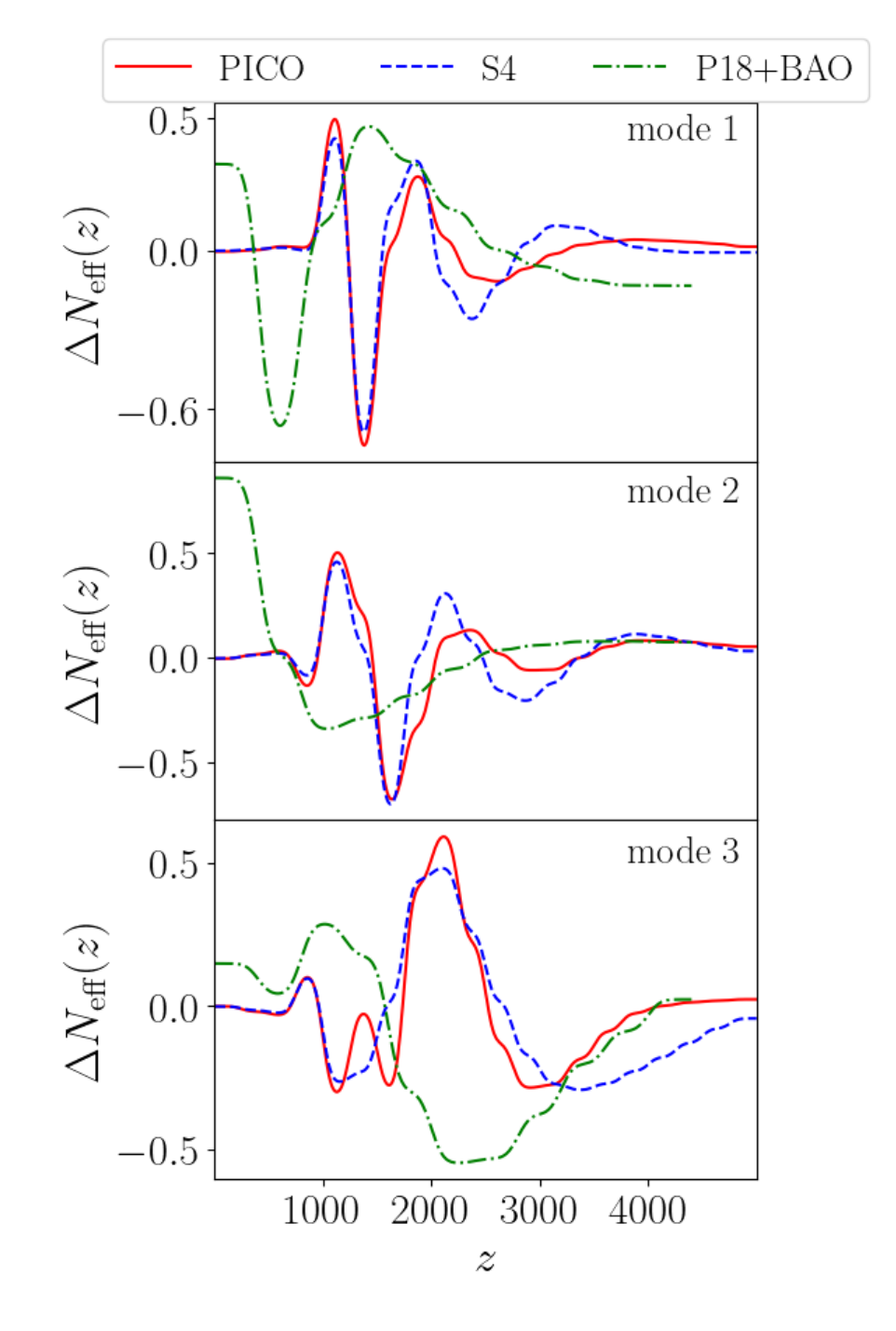}
    \caption{The first three eigenmodes constructed for PICO (solid red line), S4 (dashed blue line) and P18+BAO (dash-dotted green line).}
    \label{fig:picos4-p18bao-modes}
\end{figure}

\begin{figure}
    \centering
        \includegraphics[width=0.46\textwidth]{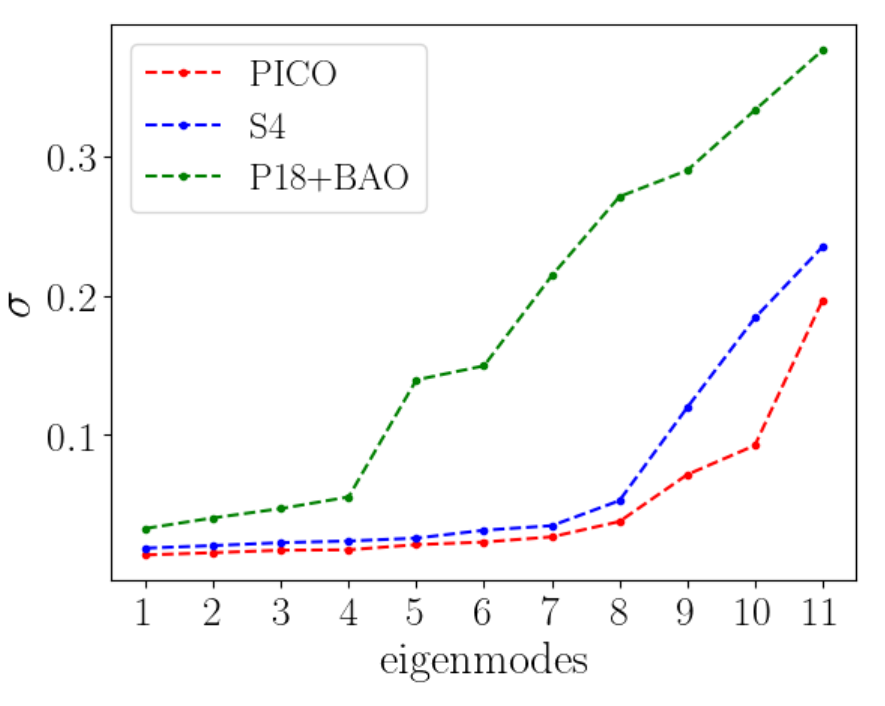} 
    \caption{Estimated errors of the eigenmodes constructed for PICO (red line), S4 (blue line) and P18+BAO (green line).}
    \label{fig:picos4_mode_errors}
\end{figure}

In this work, we use as our datasets the power spectra of CMB temperature and polarization anisotropies and CMB lensing as measured by \textit{Planck} $2018$ \citep{pl18,Planck:2018lbu}, referred to as  P18, in combination with the 
measurements of the Baryonic Acoustic Oscillations, BAO \citep{beutler20116df,Ross:2014qpa}. We also make forecasts for the futuristic CMB measurements by a PICO-like \citep{pico} and a CMB-S4-like experiment \citep{CMB-S4}, referred to as S4 throughout this work.  PICO and S4, with greatly enhanced sensitivity (equivalent to a few thousand \textit{Planck} missions) and higher angular resolution ($\ell_{\rm max}=4000$ and $3000$, respectively) will map the CMB sky (with $f_{\rm sky}=0.75$ and $0.4$, respectively) deep into its damping tail, where $\Delta N_{\rm eff}$ is expected to have the largest impact. The overall performance of the two experiments is similar in exploring the $ N_{\rm eff}(z)$ parameter space, in particular for the best constrained eigenmodes, as is evident from Figures~\ref{fig:picos4-p18bao-modes} and ~\ref{fig:picos4_mode_errors}.
Current forecasts from S4 and PICO constrain $|\Delta N_{\rm eff}|<0.06\  (95\%)$ \citep{pico,CMB-S4} which is a factor of $\sim 6$ improvement over
\textit{Planck} with $|\Delta N_{\rm eff}| < 0.37\ (95\%)$\citep{pl18}.

In the simulations, the $N_{\rm eff}$ history is assumed to agree with the standard model, thus the fiducial values of the bin amplitudes are $\nu _i^{\rm fid}=0$ for $i=1,..., n_{\rm b}$. The standard parameter values are chosen to agree with their P18 measurements in the $\Lambda$CDM scenario. We should note that neutrinos are assumed to be massless in this work. 
However, this is not expected to change the results due to the little correlation of neutrino mass and $N_{\rm eff}$ \citep{pl18}.  
In the analysis, we also assume that helium abundance $Y_{\rm p}$ is known and fixed. However, given the significant correlation between $Y_{\rm p}$ and $N_{\rm eff}$ and the weakened $N_{\rm eff}$ constraint in the presence of $Y_{\rm p}$ by a factor of $\sim 2$ \citep{pl18}, we expect  marginalization over $Y_{\rm p}$ to have noticeable impact on our results.
% \textcolor{blue}{is expected to weaken the constraints on $N_{\rm eff}$ of the by a factor of 2 (see  \citep{smith2012constraints,Nollett:2011aa,mangano2011robust,Coc:2014oia,Berlin:2019pbq,Lague:2019yvs,AristizabalSierra:2023bah}). }
\begin{table*}
\centering
\begin{tabular}{l l l l l}
%\multicolumn{3}{c}{}\\ \hline \hline
 Parameter & PICO  & S4 & P18+BAO & P18+BAO ($\Lambda \rm{CDM}$) \\
\hline
$\Omega _bh^2$ &  $0.02237 \pm 0.00003$ & $0.02236\pm 0.00004$ & $0.02248 \pm 0.00018$& $0.02242 \pm 0.00014$
\\
$\Omega _ch^2$ &  $0.1199 \pm 0.0005$ & $0.1193\pm 0.0006$ & $0.1176 \pm 0.0012$ & $0.1193 \pm 0.0009$
\\
$H_0$ & $67.95 \pm 0.18$  & $68.10 \pm 0.22$ & $68.71 \pm 0.44$ & $67.66 \pm 0.42$
\\
$n_s$ & $0.9653 \pm 0.0033$ & $0.9677 \pm 0.0026$ & $0.9725 \pm 0.0054$ & $0.9656 \pm 0.0038$
\\
$\ln (10^{10} A_s)$ & $3.044 \pm 0.003$ & $3.043 \pm 0.005$ & $3.047 \pm 0.013$ & $3.047 \pm 0.014$
\\
%$\sigma _8$ & $0.8245 \pm 0.0014$ & $0.8226 \pm 0.0018$ & $0.8206 \pm 0.0058$ & $0.8102 \pm 0.0060$
%\\
\end{tabular}
\caption{Mean and estimated errors of standard parameters for the various data combinations.}\label{table:picos4p18bao_std_mean}
\end{table*}
\begin{table}
	\centering  
	\begin{tabular}{cccccccc}
		\noalign{\smallskip}
		&${\tilde{\nu}}_1$ & ${\tilde{\nu}}_2$  &${\tilde {\nu}}_3$ \\
		\hline 
		\noalign{\smallskip}		
		P18+BAO  &$-0.008\pm  0.03 $&$-0.02\pm  0.04 $&$-0.007\pm  0.05 $&\\ \hline
		% $0.4\pm  0.6 $&	\\	\hline 
	\end{tabular}
	\caption{The amplitudes and errors of  $\Delta N_{\rm{eff}}$ eigenmodes for the P18+BAO dataset.}
	\label{tab:p18-modes}
\end{table}
%------------------------------------------------------------------
%-------------------------------------------------------------------
\section{Results}\label{sec:results}
We explore the history of $N_{\rm eff}$ in this work by taking $n_{\rm b}=11$ when analyzing P18+BAO data and  $n_{\rm b}=20$ for the PICO and S4-like cases. The bins  are linearly spaced in the redshift range $\{0, 4400\}$ for P18+BAO and  $\{0, 5000\}$ for PICO and S4 cases. 
 The results are marginalized over nuisance parameters for the P18+BAO case.
%--------------------------------------------------------------------
\begin{figure*}
    \centering
        \includegraphics[width=0.46\linewidth]{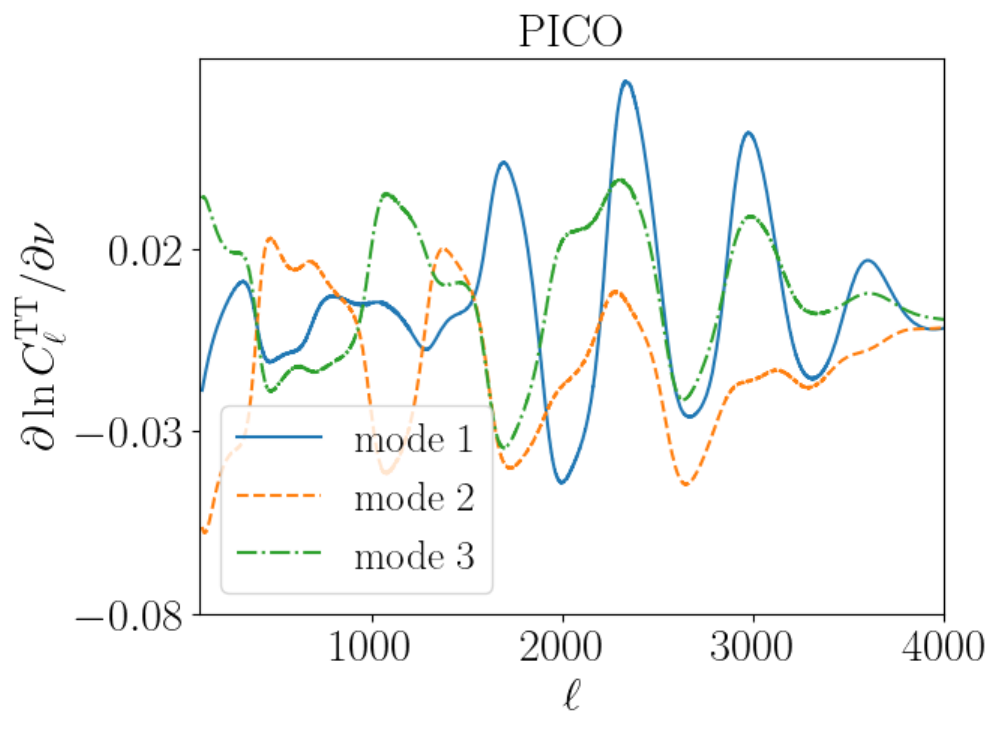}
        \includegraphics[width=0.46\linewidth]{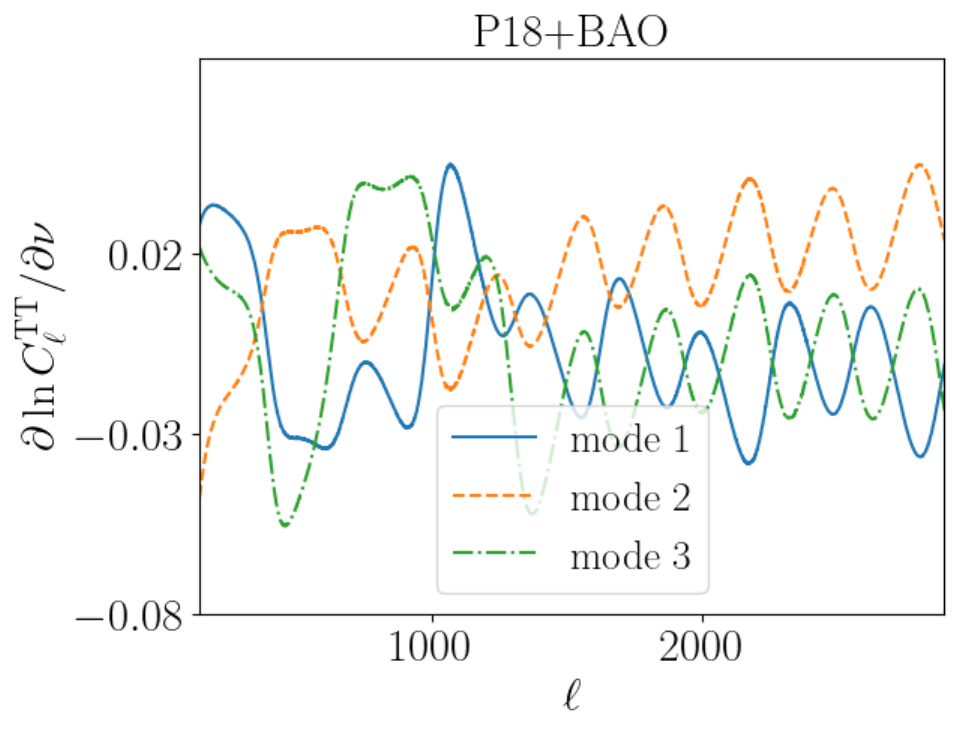}
        \caption{The sensitivity of $C^{\rm TT}_{\ell}$ to $\Delta N_{\rm eff}$ for the first three  eigenmodes for PICO (left) and P18+BAO (right).}
        \label{fig:deriv}
\end{figure*}
%-------------------------------------------------------------------
Figure~\ref{fig:pico-s4-p18-bao} shows the reconstruction of $\Delta N_{\rm eff}(z)$  from the bin amplitudes as measured  by PICO, S4 and P18+BAO. 
In PICO and S4 cases, the bins in the range 
$\{500,2000\}$ are found to have the lowest errors, while $\Delta N_{\rm eff}$ at low and high redshift ranges (with $z$ less than a few hundred and higher than 2500) is practically unconstrained. 
This could be expected from what we found in Figure~\ref{fig:bin_sensitivity}, where the orange curves, corresponding to the redshift range of $z\in \{1000,1250\}$, exhibit noticeably  large responses to changes in $N_{\rm eff}(z)$, compared to the two cases at higher and lower redshift ranges. This clearly demonstrates that CMB data are most sensitive to variation in radiation density around the recombination epoch.
 The results for P18+BAO are consistent with $\Delta N_{\rm eff}=0$. However, at lower redshifts, the bin amplitudes are found to deviate from zero. Although statistically insignificant, the deviation can raise interest, in particular if it persists with future higher resolution data.

To illustrate the correlation between the bin amplitudes, Figure~\ref{fig:corrpico} shows the correlation matrices for PICO, S4 and P18+BAO marginalized over cosmological and nuisance parameters. 
As expected, bin correlations are largest between  neighboring bins. 
In Figure~\ref{fig:pico-s4-p18-bao}, it is evident that the $1\sigma$ errors at lower redshifts are smaller for P18+BAO compared to PICO and S4. This apparent discrepancy can be attributed to the differing number of   $N_{\rm eff}(z)$ bins in these datasets. The more bins used in the case of PICO and S4 results in larger errors due to the correlation among neighboring bins. On the other hand, this enhancement in the number of parameters allows for a more detailed analysis of potential deviations from the standard $N_{\rm eff}(z)$.
The relatively high sensitivity of PICO and S4 to changes in $N_{\rm eff}(z)$  at intermediate redshifts (evident from the $C_\ell$ response in Figure~\ref{fig:bin_sensitivity} and tight $N_{\rm eff}$ constraints at those bins in Figure~\ref{fig:pico-s4-p18-bao}) is expected to lead to eigenmodes that can best probe patterns in $N_{\rm eff}$ history around those redshifts. These features are observed 
in the eigenmodes of PICO and S4 illustrated in Figure~\ref{fig:picos4-p18bao-modes}.
A similar trend is evident for P18+BAO, but with the certain  patterns of the eigenmodes at lower redshifts, where the constraints are tightest.
Moreover, the analysis presented in Figure~\ref{fig:bin_sensitivity} reveals that the majority of information from intermediate redshifts resides in higher multipoles. Therefore, due to  limited angular resolution ($\ell_{\rm max}\approx 2500$), \textit{Planck}'s sensitivity to intermediate redshifts is restricted which leads to relatively large errors.
Table~\ref{table:picos4p18bao_std_mean} presents the measurements of the standard parameters for the various datasets used in this work, marginalized 
over  the $\Delta N_{\rm eff}$ parameters.
For comparison, we have also included the P18+BAO measurements of the standard parameters within the $\Lambda$CDM framework \citep{pl18}. 
The parameter measurements are found to be consistent in the  $\Lambda$CDM scenario with the case with eleven $N_{\rm eff}$ parameters included in the 
analysis, and the discrepancies are less than $2\sigma$ for most of the parameters, except for $H_0$ where the analysis including the $N_{\rm eff}$ degrees of freedom yields a $2\sigma$ higher value for the Hubble constant.

\begin{figure}
    \centering
        \includegraphics[width=1\linewidth]{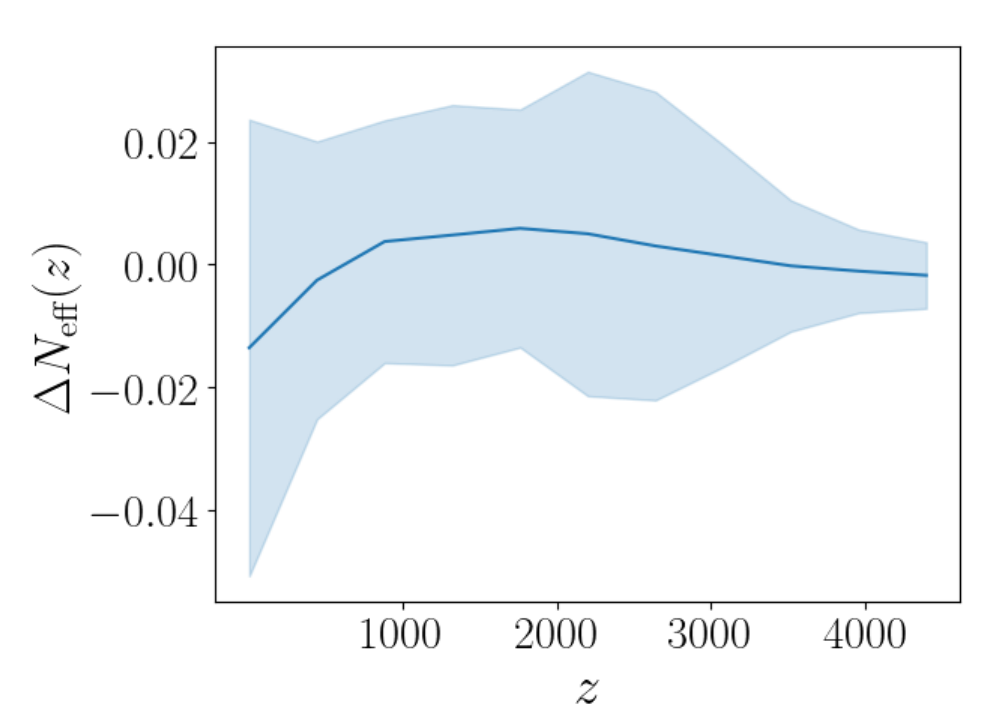}
    \caption{Reconstructed $\Delta N_{\rm eff}$ history with the first three eigenmodes for P18+BAO case as basis functions. }
    \label{fig:pca-reconst}
\end{figure}

In order to obtain the best constrainable patterns of the $N_{\rm eff}$ history, one would need to eigen-decompose the covariance matrix of the bin amplitudes. 
Using the $\nu_i$ covariance matrices found by the  post-processing of the CosmoMC Markov chains, we find the eigenmodes for  PICO, S4 and P18+BAO, along with the estimated uncertainties in their measurements.  
The first three modes are shown in Figure~\ref{fig:picos4-p18bao-modes}, and 
the mode uncertainties  are plotted in Figure~\ref{fig:picos4_mode_errors}.
Figure~\ref{fig:deriv} illustrates the sensitivity of $C_{\ell}$ to changes in $N_{\rm eff}(z)$ in the form of these eigenmodes for PICO and P18+BAO. 
The $C_{\ell}$ sensitivity to S4 modes is not plotted here as their modes are very similar to PICO (Figure~\ref{fig:picos4-p18bao-modes}). 
For the P18+BAO data, the means for the amplitudes of the first three eigenmodes are also calculated and presented
in Table~\ref{tab:p18-modes}. The results are consistent with the standard
model, and no deviation is observed, as it is also clear
from the reconstructed $N_{\rm eff}$ history using the three $\Delta N_{\rm eff}(z)$ eigenmodes of the P18+BAO
case as the basis functions (Figure~\ref{fig:pca-reconst}). The analysis makes no statement about other possible patterns of $\Delta N_{\rm eff}(z)$, orthogonal to the first three eigenmodes. Constraining those features requires higher eigenmodes to be included in the analysis. This inclusion does not  affect the measurement of the first three modes, as the modes are, by construction, uncorrelated (this should be contrasted with the $N_{\rm eff}$  reconstruction of Figure~\ref{fig:pico-s4-p18-bao}, based on top-hats where the bins had significant correlations). The current reconstruction of Figure~\ref{fig:pca-reconst} is data-driven. To compare this result with theoretical models, one should project the $\Delta N_{\rm eff}(z)$ prediction from the theory of interest onto the first three modes. If the sum of the projections goes beyond the shaded region allowed by data, the model would be observationally disfavored.
%-----------------------
%-------------------------------------------------------------------
\section{Discussion}\label{sec:conclusion}
The cosmic neutrino background, which is thermally produced in the early Universe, can be detected indirectly through its impact on CMB anisotropies. 
Any deviation of the density of this background, parameterized by $N_{\rm eff}$, from the prediction of standard model would have profound implications for 
particle physics. Future surveys, with their higher resolution observations, have the potential to more precisely constrain the contribution of relativistic species at different redshift intervals. In this work we explored the redshift dependence of $N_{\rm eff}$ in a model-independent way, investigating possible deviations 
of $N_{\rm eff}$ from the standard value of $3.044$ in terms of top-hat functions and measuring their amplitudes.
We also used the covariance matrix of the bin amplitudes to generate eigenmodes of perturbations in the $N_{\rm eff}$ history.
For the case with P18+BAO data, we used the first three eigenmodes (with the lowest errors) to reconstruct the history of $\Delta N_{\rm eff}$ in the redshift range $[0,4400]$ and found no significant deviation from the standard scenario.
Perturbations above this redshift would  practically leave no imprints on CMB. 
The inclusion of  $N_{\rm eff}$ perturbations in the analysis slightly impacted the measurement of certain other parameters, with $H_0$ enjoying the greatest ($\sim 2\sigma$) increase, from $67.66 \pm 0.42$ to $68.71 \pm 0.44$. The change is in the desired direction towards reducing the Hubble tension \citep{Riess:2021jrx,DiValentino:2021izs,Abdalla:2022yfr}. However, this result is accompanied by an increase in  $\sigma_8$ (from $0.810 \pm 0.006$ in $\Lambda$CDM scenario to  $0.821 \pm 0.006$ for the case with $\Delta N_{\rm eff}$ degrees of freedom) which implies a potential increase in the tension with data from weak lensing surveys \citep{DiValentino:2020vvd}. 
We have also reconstructed the eigenmodes for future PICO-like and CMB-S4-like surveys and found that their forecasted errors are substantially lower than the errors in the mode hierarchy for the current data. With their higher resolution and the potential to explore narrower redshift bins, one expects tighter bounds on the possible models giving   contributions to the  dark radiation component in the budget of the Universe at different epochs.

\textit{Acknowledgements}. Part of the numerical computation of this work was carried out on the computing cluster of the Canadian Institute for Theoretical Astrophysics (CITA), University of Toronto. EDV is supported by a Royal Society Dorothy Hodgkin Research Fellowship. 
This article is based upon work from COST Action CA21136 Addressing observational tensions in cosmology with systematics and fundamental physics (CosmoVerse) supported by COST (European Cooperation in Science and Technology). 
We acknowledge IT Services at The University of Sheffield for the provision of services for High Performance Computing.
This work has been supported by the Spanish MCIN/AEI/10.13039/501100011033 grants PID2020-113644GB-I00 and by the European ITN project HIDDeN (H2020-MSCA-ITN-2019/860881-HIDDeN) and SE project ASYMMETRY (HORIZON-MSCA-2021-SE-01/101086085-ASYMMETRY) and well as by the Generalitat Valenciana grants PROMETEO/2019/083 and CIPROM/2022/69. OM acknowledges the financial support from the MCIU with funding from the European Union NextGenerationEU (PRTR-C17.I01) and Generalitat Valenciana (ASFAE/2022/020).
\bibliography{pcaneff}% Produces the bibliography via BibTeX.

%merlin.mbs apsrev4-1.bst 2010-07-25 4.21a (PWD, AO, DPC) hacked
%Control: key (0)
%Control: author (72) initials jnrlst
%Control: editor formatted (1) identically to author
%Control: production of article title (-1) disabled
%Control: page (0) single
%Control: year (1) truncated
%Control: production of eprint (0) enabled
\begin{thebibliography}{39}%
\makeatletter
\providecommand \@ifxundefined [1]{%
 \@ifx{#1\undefined}
}%
\providecommand \@ifnum [1]{%
 \ifnum #1\expandafter \@firstoftwo
 \else \expandafter \@secondoftwo
 \fi
}%
\providecommand \@ifx [1]{%
 \ifx #1\expandafter \@firstoftwo
 \else \expandafter \@secondoftwo
 \fi
}%
\providecommand \natexlab [1]{#1}%
\providecommand \enquote  [1]{``#1''}%
\providecommand \bibnamefont  [1]{#1}%
\providecommand \bibfnamefont [1]{#1}%
\providecommand \citenamefont [1]{#1}%
\providecommand \href@noop [0]{\@secondoftwo}%
\providecommand \href [0]{\begingroup \@sanitize@url \@href}%
\providecommand \@href[1]{\@@startlink{#1}\@@href}%
\providecommand \@@href[1]{\endgroup#1\@@endlink}%
\providecommand \@sanitize@url [0]{\catcode `\\12\catcode `\$12\catcode
  `\&12\catcode `\#12\catcode `\^12\catcode `\_12\catcode `\%12\relax}%
\providecommand \@@startlink[1]{}%
\providecommand \@@endlink[0]{}%
\providecommand \url  [0]{\begingroup\@sanitize@url \@url }%
\providecommand \@url [1]{\endgroup\@href {#1}{\urlprefix }}%
\providecommand \urlprefix  [0]{URL }%
\providecommand \Eprint [0]{\href }%
\providecommand \doibase [0]{http://dx.doi.org/}%
\providecommand \selectlanguage [0]{\@gobble}%
\providecommand \bibinfo  [0]{\@secondoftwo}%
\providecommand \bibfield  [0]{\@secondoftwo}%
\providecommand \translation [1]{[#1]}%
\providecommand \BibitemOpen [0]{}%
\providecommand \bibitemStop [0]{}%
\providecommand \bibitemNoStop [0]{.\EOS\space}%
\providecommand \EOS [0]{\spacefactor3000\relax}%
\providecommand \BibitemShut  [1]{\csname bibitem#1\endcsname}%
\let\auto@bib@innerbib\@empty
%</preamble>
\bibitem [{\citenamefont {Akita}\ and\ \citenamefont
  {Yamaguchi}(2020)}]{Akita:2020szl}%
  \BibitemOpen
  \bibfield  {author} {\bibinfo {author} {\bibfnamefont {K.}~\bibnamefont
  {Akita}}\ and\ \bibinfo {author} {\bibfnamefont {M.}~\bibnamefont
  {Yamaguchi}},\ }\href {\doibase 10.1088/1475-7516/2020/08/012} {\bibfield
  {journal} {\bibinfo  {journal} {JCAP}\ }\textbf {\bibinfo {volume} {08}},\
  \bibinfo {pages} {012} (\bibinfo {year} {2020})},\ \Eprint
  {http://arxiv.org/abs/2005.07047} {arXiv:2005.07047 [hep-ph]} \BibitemShut
  {NoStop}%
\bibitem [{\citenamefont {Froustey}\ \emph {et~al.}(2020)\citenamefont
  {Froustey}, \citenamefont {Pitrou},\ and\ \citenamefont
  {Volpe}}]{Froustey:2020mcq}%
  \BibitemOpen
  \bibfield  {author} {\bibinfo {author} {\bibfnamefont {J.}~\bibnamefont
  {Froustey}}, \bibinfo {author} {\bibfnamefont {C.}~\bibnamefont {Pitrou}}, \
  and\ \bibinfo {author} {\bibfnamefont {M.~C.}\ \bibnamefont {Volpe}},\ }\href
  {\doibase 10.1088/1475-7516/2020/12/015} {\bibfield  {journal} {\bibinfo
  {journal} {JCAP}\ }\textbf {\bibinfo {volume} {12}},\ \bibinfo {pages} {015}
  (\bibinfo {year} {2020})},\ \Eprint {http://arxiv.org/abs/2008.01074}
  {arXiv:2008.01074 [hep-ph]} \BibitemShut {NoStop}%
\bibitem [{\citenamefont {Bennett}\ \emph {et~al.}(2021)\citenamefont
  {Bennett}, \citenamefont {Buldgen}, \citenamefont {De~Salas}, \citenamefont
  {Drewes}, \citenamefont {Gariazzo}, \citenamefont {Pastor},\ and\
  \citenamefont {Wong}}]{Bennett:2020zkv}%
  \BibitemOpen
  \bibfield  {author} {\bibinfo {author} {\bibfnamefont {J.~J.}\ \bibnamefont
  {Bennett}}, \bibinfo {author} {\bibfnamefont {G.}~\bibnamefont {Buldgen}},
  \bibinfo {author} {\bibfnamefont {P.~F.}\ \bibnamefont {De~Salas}}, \bibinfo
  {author} {\bibfnamefont {M.}~\bibnamefont {Drewes}}, \bibinfo {author}
  {\bibfnamefont {S.}~\bibnamefont {Gariazzo}}, \bibinfo {author}
  {\bibfnamefont {S.}~\bibnamefont {Pastor}}, \ and\ \bibinfo {author}
  {\bibfnamefont {Y.~Y.~Y.}\ \bibnamefont {Wong}},\ }\href {\doibase
  10.1088/1475-7516/2021/04/073} {\bibfield  {journal} {\bibinfo  {journal}
  {JCAP}\ }\textbf {\bibinfo {volume} {04}},\ \bibinfo {pages} {073} (\bibinfo
  {year} {2021})},\ \Eprint {http://arxiv.org/abs/2012.02726} {arXiv:2012.02726
  [hep-ph]} \BibitemShut {NoStop}%
\bibitem [{\citenamefont {Abazajian}\ \emph {et~al.}(2015)\citenamefont
  {Abazajian} \emph {et~al.}}]{AbazajianCarlstromLee:2013}%
  \BibitemOpen
  \bibfield  {author} {\bibinfo {author} {\bibfnamefont {K.~N.}\ \bibnamefont
  {Abazajian}} \emph {et~al.} (\bibinfo {collaboration} {Topical Conveners:
  K.N. Abazajian, J.E. Carlstrom, A.T. Lee}),\ }\href {\doibase
  10.1016/j.astropartphys.2014.05.014} {\bibfield  {journal} {\bibinfo
  {journal} {Astropart. Phys.}\ }\textbf {\bibinfo {volume} {63}},\ \bibinfo
  {pages} {66} (\bibinfo {year} {2015})},\ \Eprint
  {http://arxiv.org/abs/1309.5383} {arXiv:1309.5383 [astro-ph.CO]} \BibitemShut
  {NoStop}%
\bibitem [{\citenamefont {Calabrese}\ \emph {et~al.}(2011)\citenamefont
  {Calabrese}, \citenamefont {Huterer}, \citenamefont {Linder}, \citenamefont
  {Melchiorri},\ and\ \citenamefont {Pagano}}]{calabrese2011limits}%
  \BibitemOpen
  \bibfield  {author} {\bibinfo {author} {\bibfnamefont {E.}~\bibnamefont
  {Calabrese}}, \bibinfo {author} {\bibfnamefont {D.}~\bibnamefont {Huterer}},
  \bibinfo {author} {\bibfnamefont {E.~V.}\ \bibnamefont {Linder}}, \bibinfo
  {author} {\bibfnamefont {A.}~\bibnamefont {Melchiorri}}, \ and\ \bibinfo
  {author} {\bibfnamefont {L.}~\bibnamefont {Pagano}},\ }\href@noop {}
  {\bibfield  {journal} {\bibinfo  {journal} {Physical Review D}\ }\textbf
  {\bibinfo {volume} {83}},\ \bibinfo {pages} {123504} (\bibinfo {year}
  {2011})}\BibitemShut {NoStop}%
\bibitem [{\citenamefont {Archidiacono}\ \emph {et~al.}(2011)\citenamefont
  {Archidiacono}, \citenamefont {Calabrese},\ and\ \citenamefont
  {Melchiorri}}]{Archidiacono:2011gq}%
  \BibitemOpen
  \bibfield  {author} {\bibinfo {author} {\bibfnamefont {M.}~\bibnamefont
  {Archidiacono}}, \bibinfo {author} {\bibfnamefont {E.}~\bibnamefont
  {Calabrese}}, \ and\ \bibinfo {author} {\bibfnamefont {A.}~\bibnamefont
  {Melchiorri}},\ }\href {\doibase 10.1103/PhysRevD.84.123008} {\bibfield
  {journal} {\bibinfo  {journal} {Phys. Rev. D}\ }\textbf {\bibinfo {volume}
  {84}},\ \bibinfo {pages} {123008} (\bibinfo {year} {2011})},\ \Eprint
  {http://arxiv.org/abs/1109.2767} {arXiv:1109.2767 [astro-ph.CO]} \BibitemShut
  {NoStop}%
\bibitem [{\citenamefont {Joudaki}\ \emph {et~al.}(2013)\citenamefont
  {Joudaki}, \citenamefont {Abazajian},\ and\ \citenamefont
  {Kaplinghat}}]{Joudaki:2012uk}%
  \BibitemOpen
  \bibfield  {author} {\bibinfo {author} {\bibfnamefont {S.}~\bibnamefont
  {Joudaki}}, \bibinfo {author} {\bibfnamefont {K.~N.}\ \bibnamefont
  {Abazajian}}, \ and\ \bibinfo {author} {\bibfnamefont {M.}~\bibnamefont
  {Kaplinghat}},\ }\href {\doibase 10.1103/PhysRevD.87.065003} {\bibfield
  {journal} {\bibinfo  {journal} {Phys. Rev. D}\ }\textbf {\bibinfo {volume}
  {87}},\ \bibinfo {pages} {065003} (\bibinfo {year} {2013})},\ \Eprint
  {http://arxiv.org/abs/1208.4354} {arXiv:1208.4354 [astro-ph.CO]} \BibitemShut
  {NoStop}%
\bibitem [{\citenamefont {Jacques}\ \emph {et~al.}(2013)\citenamefont
  {Jacques}, \citenamefont {Krauss},\ and\ \citenamefont
  {Lunardini}}]{Jacques:2013xr}%
  \BibitemOpen
  \bibfield  {author} {\bibinfo {author} {\bibfnamefont {T.~D.}\ \bibnamefont
  {Jacques}}, \bibinfo {author} {\bibfnamefont {L.~M.}\ \bibnamefont {Krauss}},
  \ and\ \bibinfo {author} {\bibfnamefont {C.}~\bibnamefont {Lunardini}},\
  }\href {\doibase 10.1103/PhysRevD.87.083515} {\bibfield  {journal} {\bibinfo
  {journal} {Phys. Rev. D}\ }\textbf {\bibinfo {volume} {87}},\ \bibinfo
  {pages} {083515} (\bibinfo {year} {2013})},\ \bibinfo {note} {[ Erratum:
  Phys.Rev.D 88, 109901 (2013)]},\ \Eprint {http://arxiv.org/abs/1301.3119}
  {arXiv:1301.3119 [astro-ph.CO]} \BibitemShut {NoStop}%
\bibitem [{\citenamefont {Gariazzo}\ \emph {et~al.}(2016)\citenamefont
  {Gariazzo}, \citenamefont {Giunti}, \citenamefont {Laveder}, \citenamefont
  {Li},\ and\ \citenamefont {Zavanin}}]{Gariazzo:2015rra}%
  \BibitemOpen
  \bibfield  {author} {\bibinfo {author} {\bibfnamefont {S.}~\bibnamefont
  {Gariazzo}}, \bibinfo {author} {\bibfnamefont {C.}~\bibnamefont {Giunti}},
  \bibinfo {author} {\bibfnamefont {M.}~\bibnamefont {Laveder}}, \bibinfo
  {author} {\bibfnamefont {Y.~F.}\ \bibnamefont {Li}}, \ and\ \bibinfo {author}
  {\bibfnamefont {E.~M.}\ \bibnamefont {Zavanin}},\ }\href {\doibase
  10.1088/0954-3899/43/3/033001} {\bibfield  {journal} {\bibinfo  {journal} {J.
  Phys. G}\ }\textbf {\bibinfo {volume} {43}},\ \bibinfo {pages} {033001}
  (\bibinfo {year} {2016})},\ \Eprint {http://arxiv.org/abs/1507.08204}
  {arXiv:1507.08204 [hep-ph]} \BibitemShut {NoStop}%
\bibitem [{\citenamefont {Archidiacono}\ and\ \citenamefont
  {Gariazzo}(2022)}]{Archidiacono:2022ich}%
  \BibitemOpen
  \bibfield  {author} {\bibinfo {author} {\bibfnamefont {M.}~\bibnamefont
  {Archidiacono}}\ and\ \bibinfo {author} {\bibfnamefont {S.}~\bibnamefont
  {Gariazzo}},\ }\href {\doibase 10.3390/universe8030175} {\bibfield  {journal}
  {\bibinfo  {journal} {Universe}\ }\textbf {\bibinfo {volume} {8}},\ \bibinfo
  {pages} {175} (\bibinfo {year} {2022})},\ \Eprint
  {http://arxiv.org/abs/2201.10319} {arXiv:2201.10319 [hep-ph]} \BibitemShut
  {NoStop}%
\bibitem [{\citenamefont {Arbey}\ \emph {et~al.}(2021)\citenamefont {Arbey},
  \citenamefont {Auffinger}, \citenamefont {Sandick}, \citenamefont {Shams
  Es~Haghi},\ and\ \citenamefont {Sinha}}]{Arbey:2021ysg}%
  \BibitemOpen
  \bibfield  {author} {\bibinfo {author} {\bibfnamefont {A.}~\bibnamefont
  {Arbey}}, \bibinfo {author} {\bibfnamefont {J.}~\bibnamefont {Auffinger}},
  \bibinfo {author} {\bibfnamefont {P.}~\bibnamefont {Sandick}}, \bibinfo
  {author} {\bibfnamefont {B.}~\bibnamefont {Shams Es~Haghi}}, \ and\ \bibinfo
  {author} {\bibfnamefont {K.}~\bibnamefont {Sinha}},\ }\href {\doibase
  10.1103/PhysRevD.103.123549} {\bibfield  {journal} {\bibinfo  {journal}
  {Phys. Rev. D}\ }\textbf {\bibinfo {volume} {103}},\ \bibinfo {pages}
  {123549} (\bibinfo {year} {2021})},\ \Eprint
  {http://arxiv.org/abs/2104.04051} {arXiv:2104.04051 [astro-ph.CO]}
  \BibitemShut {NoStop}%
\bibitem [{\citenamefont {Abazajian}\ and\ \citenamefont
  {Heeck}(2019)}]{Abazajian:2019oqj}%
  \BibitemOpen
  \bibfield  {author} {\bibinfo {author} {\bibfnamefont {K.~N.}\ \bibnamefont
  {Abazajian}}\ and\ \bibinfo {author} {\bibfnamefont {J.}~\bibnamefont
  {Heeck}},\ }\href {\doibase 10.1103/PhysRevD.100.075027} {\bibfield
  {journal} {\bibinfo  {journal} {Phys. Rev. D}\ }\textbf {\bibinfo {volume}
  {100}},\ \bibinfo {pages} {075027} (\bibinfo {year} {2019})},\ \Eprint
  {http://arxiv.org/abs/1908.03286} {arXiv:1908.03286 [hep-ph]} \BibitemShut
  {NoStop}%
\bibitem [{\citenamefont {Luo}\ \emph {et~al.}(2020)\citenamefont {Luo},
  \citenamefont {Rodejohann},\ and\ \citenamefont {Xu}}]{Luo:2020sho}%
  \BibitemOpen
  \bibfield  {author} {\bibinfo {author} {\bibfnamefont {X.}~\bibnamefont
  {Luo}}, \bibinfo {author} {\bibfnamefont {W.}~\bibnamefont {Rodejohann}}, \
  and\ \bibinfo {author} {\bibfnamefont {X.-J.}\ \bibnamefont {Xu}},\ }\href
  {\doibase 10.1088/1475-7516/2020/06/058} {\bibfield  {journal} {\bibinfo
  {journal} {JCAP}\ }\textbf {\bibinfo {volume} {06}},\ \bibinfo {pages} {058}
  (\bibinfo {year} {2020})},\ \Eprint {http://arxiv.org/abs/2005.01629}
  {arXiv:2005.01629 [hep-ph]} \BibitemShut {NoStop}%
\bibitem [{\citenamefont {Conlon}\ and\ \citenamefont
  {Marsh}(2013)}]{Conlon:2013isa}%
  \BibitemOpen
  \bibfield  {author} {\bibinfo {author} {\bibfnamefont {J.~P.}\ \bibnamefont
  {Conlon}}\ and\ \bibinfo {author} {\bibfnamefont {M.~C.~D.}\ \bibnamefont
  {Marsh}},\ }\href {\doibase 10.1007/JHEP10(2013)214} {\bibfield  {journal}
  {\bibinfo  {journal} {JHEP}\ }\textbf {\bibinfo {volume} {10}},\ \bibinfo
  {pages} {214} (\bibinfo {year} {2013})},\ \Eprint
  {http://arxiv.org/abs/1304.1804} {arXiv:1304.1804 [hep-ph]} \BibitemShut
  {NoStop}%
\bibitem [{\citenamefont {Baumann}\ \emph {et~al.}(2016)\citenamefont
  {Baumann}, \citenamefont {Green},\ and\ \citenamefont
  {Wallisch}}]{Baumann:2016wac}%
  \BibitemOpen
  \bibfield  {author} {\bibinfo {author} {\bibfnamefont {D.}~\bibnamefont
  {Baumann}}, \bibinfo {author} {\bibfnamefont {D.}~\bibnamefont {Green}}, \
  and\ \bibinfo {author} {\bibfnamefont {B.}~\bibnamefont {Wallisch}},\ }\href
  {\doibase 10.1103/PhysRevLett.117.171301} {\bibfield  {journal} {\bibinfo
  {journal} {Phys. Rev. Lett.}\ }\textbf {\bibinfo {volume} {117}},\ \bibinfo
  {pages} {171301} (\bibinfo {year} {2016})},\ \Eprint
  {http://arxiv.org/abs/1604.08614} {arXiv:1604.08614 [astro-ph.CO]}
  \BibitemShut {NoStop}%
\bibitem [{\citenamefont {Gelmini}\ \emph {et~al.}(2004)\citenamefont
  {Gelmini}, \citenamefont {Palomares-Ruiz},\ and\ \citenamefont
  {Pascoli}}]{Gelmini:2004ah}%
  \BibitemOpen
  \bibfield  {author} {\bibinfo {author} {\bibfnamefont {G.}~\bibnamefont
  {Gelmini}}, \bibinfo {author} {\bibfnamefont {S.}~\bibnamefont
  {Palomares-Ruiz}}, \ and\ \bibinfo {author} {\bibfnamefont {S.}~\bibnamefont
  {Pascoli}},\ }\href {\doibase 10.1103/PhysRevLett.93.081302} {\bibfield
  {journal} {\bibinfo  {journal} {Phys. Rev. Lett.}\ }\textbf {\bibinfo
  {volume} {93}},\ \bibinfo {pages} {081302} (\bibinfo {year} {2004})},\
  \Eprint {http://arxiv.org/abs/astro-ph/0403323} {arXiv:astro-ph/0403323}
  \BibitemShut {NoStop}%
\bibitem [{\citenamefont {Finkbeiner}\ \emph {et~al.}(2012)\citenamefont
  {Finkbeiner}, \citenamefont {Galli}, \citenamefont {Lin},\ and\ \citenamefont
  {Slatyer}}]{finkbeiner2012searching}%
  \BibitemOpen
  \bibfield  {author} {\bibinfo {author} {\bibfnamefont {D.~P.}\ \bibnamefont
  {Finkbeiner}}, \bibinfo {author} {\bibfnamefont {S.}~\bibnamefont {Galli}},
  \bibinfo {author} {\bibfnamefont {T.}~\bibnamefont {Lin}}, \ and\ \bibinfo
  {author} {\bibfnamefont {T.~R.}\ \bibnamefont {Slatyer}},\ }\href@noop {}
  {\bibfield  {journal} {\bibinfo  {journal} {Physical Review D—Particles,
  Fields, Gravitation, and Cosmology}\ }\textbf {\bibinfo {volume} {85}},\
  \bibinfo {pages} {043522} (\bibinfo {year} {2012})}\BibitemShut {NoStop}%
\bibitem [{\citenamefont {Poulin}\ \emph {et~al.}(2017)\citenamefont {Poulin},
  \citenamefont {Lesgourgues},\ and\ \citenamefont {Serpico}}]{Poulin:2016anj}%
  \BibitemOpen
  \bibfield  {author} {\bibinfo {author} {\bibfnamefont {V.}~\bibnamefont
  {Poulin}}, \bibinfo {author} {\bibfnamefont {J.}~\bibnamefont {Lesgourgues}},
  \ and\ \bibinfo {author} {\bibfnamefont {P.~D.}\ \bibnamefont {Serpico}},\
  }\href {\doibase 10.1088/1475-7516/2017/03/043} {\bibfield  {journal}
  {\bibinfo  {journal} {JCAP}\ }\textbf {\bibinfo {volume} {03}},\ \bibinfo
  {pages} {043} (\bibinfo {year} {2017})},\ \Eprint
  {http://arxiv.org/abs/1610.10051} {arXiv:1610.10051 [astro-ph.CO]}
  \BibitemShut {NoStop}%
\bibitem [{\citenamefont {Freese}\ \emph {et~al.}(2018)\citenamefont {Freese},
  \citenamefont {Sfakianakis}, \citenamefont {Stengel},\ and\ \citenamefont
  {Visinelli}}]{Freese:2017ace}%
  \BibitemOpen
  \bibfield  {author} {\bibinfo {author} {\bibfnamefont {K.}~\bibnamefont
  {Freese}}, \bibinfo {author} {\bibfnamefont {E.~I.}\ \bibnamefont
  {Sfakianakis}}, \bibinfo {author} {\bibfnamefont {P.}~\bibnamefont
  {Stengel}}, \ and\ \bibinfo {author} {\bibfnamefont {L.}~\bibnamefont
  {Visinelli}},\ }\href {\doibase 10.1088/1475-7516/2018/05/067} {\bibfield
  {journal} {\bibinfo  {journal} {JCAP}\ }\textbf {\bibinfo {volume} {05}},\
  \bibinfo {pages} {067} (\bibinfo {year} {2018})},\ \Eprint
  {http://arxiv.org/abs/1712.03791} {arXiv:1712.03791 [hep-ph]} \BibitemShut
  {NoStop}%
\bibitem [{\citenamefont {Aghanim}\ \emph
  {et~al.}(2020{\natexlab{a}})\citenamefont {Aghanim} \emph {et~al.}}]{pl18}%
  \BibitemOpen
  \bibfield  {author} {\bibinfo {author} {\bibfnamefont {N.}~\bibnamefont
  {Aghanim}} \emph {et~al.} (\bibinfo {collaboration} {Planck}),\ }\href
  {\doibase 10.1051/0004-6361/201833910} {\bibfield  {journal} {\bibinfo
  {journal} {Astron. Astrophys.}\ }\textbf {\bibinfo {volume} {641}},\ \bibinfo
  {pages} {A6} (\bibinfo {year} {2020}{\natexlab{a}})},\ \bibinfo {note}
  {[Erratum: Astron.Astrophys. 652, C4 (2021)]},\ \Eprint
  {http://arxiv.org/abs/1807.06209} {arXiv:1807.06209 [astro-ph.CO]}
  \BibitemShut {NoStop}%
\bibitem [{\citenamefont {Hanany}\ \emph {et~al.}(2019)\citenamefont {Hanany}
  \emph {et~al.}}]{pico}%
  \BibitemOpen
  \bibfield  {author} {\bibinfo {author} {\bibfnamefont {S.}~\bibnamefont
  {Hanany}} \emph {et~al.} (\bibinfo {collaboration} {NASA PICO}),\ }\href@noop
  {} {\  (\bibinfo {year} {2019})},\ \Eprint {http://arxiv.org/abs/1902.10541}
  {arXiv:1902.10541 [astro-ph.IM]} \BibitemShut {NoStop}%
\bibitem [{\citenamefont {Abazajian}\ \emph {et~al.}(2016)\citenamefont
  {Abazajian} \emph {et~al.}}]{CMB-S4}%
  \BibitemOpen
  \bibfield  {author} {\bibinfo {author} {\bibfnamefont {K.~N.}\ \bibnamefont
  {Abazajian}} \emph {et~al.} (\bibinfo {collaboration} {CMB-S4}),\ }\href@noop
  {} {\  (\bibinfo {year} {2016})},\ \Eprint {http://arxiv.org/abs/1610.02743}
  {arXiv:1610.02743 [astro-ph.CO]} \BibitemShut {NoStop}%
\bibitem [{\citenamefont {Lorenz}\ \emph {et~al.}(2021)\citenamefont {Lorenz},
  \citenamefont {Funcke}, \citenamefont {L\"offler},\ and\ \citenamefont
  {Calabrese}}]{Lorenz:2021alz}%
  \BibitemOpen
  \bibfield  {author} {\bibinfo {author} {\bibfnamefont {C.~S.}\ \bibnamefont
  {Lorenz}}, \bibinfo {author} {\bibfnamefont {L.}~\bibnamefont {Funcke}},
  \bibinfo {author} {\bibfnamefont {M.}~\bibnamefont {L\"offler}}, \ and\
  \bibinfo {author} {\bibfnamefont {E.}~\bibnamefont {Calabrese}},\ }\href
  {\doibase 10.1103/PhysRevD.104.123518} {\bibfield  {journal} {\bibinfo
  {journal} {Phys. Rev. D}\ }\textbf {\bibinfo {volume} {104}},\ \bibinfo
  {pages} {123518} (\bibinfo {year} {2021})},\ \Eprint
  {http://arxiv.org/abs/2102.13618} {arXiv:2102.13618 [astro-ph.CO]}
  \BibitemShut {NoStop}%
\bibitem [{\citenamefont {Lewis}\ and\ \citenamefont {Bridle}(2002)}]{lewis02}%
  \BibitemOpen
  \bibfield  {author} {\bibinfo {author} {\bibfnamefont {A.}~\bibnamefont
  {Lewis}}\ and\ \bibinfo {author} {\bibfnamefont {S.}~\bibnamefont {Bridle}},\
  }\href {\doibase 10.1103/PhysRevD.66.103511} {\bibfield  {journal} {\bibinfo
  {journal} {Phys. Rev. D}\ }\textbf {\bibinfo {volume} {66}},\ \bibinfo
  {pages} {103511} (\bibinfo {year} {2002})},\ \Eprint
  {http://arxiv.org/abs/astro-ph/0205436} {arXiv:astro-ph/0205436} \BibitemShut
  {NoStop}%
\bibitem [{\citenamefont {Lewis}(2013)}]{lewis13}%
  \BibitemOpen
  \bibfield  {author} {\bibinfo {author} {\bibfnamefont {A.}~\bibnamefont
  {Lewis}},\ }\href {\doibase 10.1103/PhysRevD.87.103529} {\bibfield  {journal}
  {\bibinfo  {journal} {Phys. Rev. D}\ }\textbf {\bibinfo {volume} {87}},\
  \bibinfo {pages} {103529} (\bibinfo {year} {2013})},\ \Eprint
  {http://arxiv.org/abs/1304.4473} {arXiv:1304.4473 [astro-ph.CO]} \BibitemShut
  {NoStop}%
\bibitem [{\citenamefont {Farhang}\ \emph {et~al.}(2012)\citenamefont
  {Farhang}, \citenamefont {Bond},\ and\ \citenamefont
  {Chluba}}]{Farhang_2012}%
  \BibitemOpen
  \bibfield  {author} {\bibinfo {author} {\bibfnamefont {M.}~\bibnamefont
  {Farhang}}, \bibinfo {author} {\bibfnamefont {J.~R.}\ \bibnamefont {Bond}}, \
  and\ \bibinfo {author} {\bibfnamefont {J.}~\bibnamefont {Chluba}},\ }\href
  {\doibase 10.1088/0004-637x/752/2/88} {\bibfield  {journal} {\bibinfo
  {journal} {The Astrophysical Journal}\ }\textbf {\bibinfo {volume} {752}},\
  \bibinfo {pages} {88} (\bibinfo {year} {2012})}\BibitemShut {NoStop}%
\bibitem [{\citenamefont {Farhang}\ \emph {et~al.}(2013)\citenamefont
  {Farhang}, \citenamefont {Bond}, \citenamefont {Chluba},\ and\ \citenamefont
  {Switzer}}]{Farhang_2013}%
  \BibitemOpen
  \bibfield  {author} {\bibinfo {author} {\bibfnamefont {M.}~\bibnamefont
  {Farhang}}, \bibinfo {author} {\bibfnamefont {J.~R.}\ \bibnamefont {Bond}},
  \bibinfo {author} {\bibfnamefont {J.}~\bibnamefont {Chluba}}, \ and\ \bibinfo
  {author} {\bibfnamefont {E.~R.}\ \bibnamefont {Switzer}},\ }\href {\doibase
  10.1088/0004-637x/764/2/137} {\bibfield  {journal} {\bibinfo  {journal} {The
  Astrophysical Journal}\ }\textbf {\bibinfo {volume} {764}},\ \bibinfo {pages}
  {137} (\bibinfo {year} {2013})}\BibitemShut {NoStop}%
\bibitem [{\citenamefont {Pandolfi}\ \emph {et~al.}(2010)\citenamefont
  {Pandolfi}, \citenamefont {Giusarma}, \citenamefont {Kolb}, \citenamefont
  {Lattanzi}, \citenamefont {Melchiorri}, \citenamefont {Mena}, \citenamefont
  {Pe{\~{n} }a}, \citenamefont {Cooray},\ and\ \citenamefont
  {Serra}}]{Pandolfi_2010}%
  \BibitemOpen
  \bibfield  {author} {\bibinfo {author} {\bibfnamefont {S.}~\bibnamefont
  {Pandolfi}}, \bibinfo {author} {\bibfnamefont {E.}~\bibnamefont {Giusarma}},
  \bibinfo {author} {\bibfnamefont {E.~W.}\ \bibnamefont {Kolb}}, \bibinfo
  {author} {\bibfnamefont {M.}~\bibnamefont {Lattanzi}}, \bibinfo {author}
  {\bibfnamefont {A.}~\bibnamefont {Melchiorri}}, \bibinfo {author}
  {\bibfnamefont {O.}~\bibnamefont {Mena}}, \bibinfo {author} {\bibfnamefont
  {M.}~\bibnamefont {Pe{\~{n} }a}}, \bibinfo {author} {\bibfnamefont
  {A.}~\bibnamefont {Cooray}}, \ and\ \bibinfo {author} {\bibfnamefont
  {P.}~\bibnamefont {Serra}},\ }\href {\doibase 10.1103/physrevd.82.123527}
  {\bibfield  {journal} {\bibinfo  {journal} {Physical Review D}\ }\textbf
  {\bibinfo {volume} {82}} (\bibinfo {year} {2010}),\
  10.1103/physrevd.82.123527}\BibitemShut {NoStop}%
\bibitem [{\citenamefont {Villanueva-Domingo}\ \emph
  {et~al.}(2018)\citenamefont {Villanueva-Domingo}, \citenamefont {Gariazzo},
  \citenamefont {Gnedin},\ and\ \citenamefont
  {Mena}}]{Villanueva_Domingo_2018}%
  \BibitemOpen
  \bibfield  {author} {\bibinfo {author} {\bibfnamefont {P.}~\bibnamefont
  {Villanueva-Domingo}}, \bibinfo {author} {\bibfnamefont {S.}~\bibnamefont
  {Gariazzo}}, \bibinfo {author} {\bibfnamefont {N.~Y.}\ \bibnamefont
  {Gnedin}}, \ and\ \bibinfo {author} {\bibfnamefont {O.}~\bibnamefont
  {Mena}},\ }\href {\doibase 10.1088/1475-7516/2018/04/024} {\bibfield
  {journal} {\bibinfo  {journal} {Journal of Cosmology and Astroparticle
  Physics}\ }\textbf {\bibinfo {volume} {2018}},\ \bibinfo {pages} {024}
  (\bibinfo {year} {2018})}\BibitemShut {NoStop}%
\bibitem [{\citenamefont {Farhang}\ and\ \citenamefont
  {Sadr}(2019)}]{Farhang_2019}%
  \BibitemOpen
  \bibfield  {author} {\bibinfo {author} {\bibfnamefont {M.}~\bibnamefont
  {Farhang}}\ and\ \bibinfo {author} {\bibfnamefont {A.~V.}\ \bibnamefont
  {Sadr}},\ }\href {\doibase 10.3847/1538-4357/aaf7a1} {\bibfield  {journal}
  {\bibinfo  {journal} {The Astrophysical Journal}\ }\textbf {\bibinfo {volume}
  {871}},\ \bibinfo {pages} {139} (\bibinfo {year} {2019})}\BibitemShut
  {NoStop}%
\bibitem [{\citenamefont {Esmaeilian}\ \emph {et~al.}(2021)\citenamefont
  {Esmaeilian}, \citenamefont {Farhang},\ and\ \citenamefont
  {Khodabakhshi}}]{Esmaeilian_2021}%
  \BibitemOpen
  \bibfield  {author} {\bibinfo {author} {\bibfnamefont {M.~S.}\ \bibnamefont
  {Esmaeilian}}, \bibinfo {author} {\bibfnamefont {M.}~\bibnamefont {Farhang}},
  \ and\ \bibinfo {author} {\bibfnamefont {S.}~\bibnamefont {Khodabakhshi}},\
  }\href {\doibase 10.3847/1538-4357/abe865} {\bibfield  {journal} {\bibinfo
  {journal} {The Astrophysical Journal}\ }\textbf {\bibinfo {volume} {912}},\
  \bibinfo {pages} {104} (\bibinfo {year} {2021})}\BibitemShut {NoStop}%
\bibitem [{\citenamefont {Aghanim}\ \emph
  {et~al.}(2020{\natexlab{b}})\citenamefont {Aghanim} \emph
  {et~al.}}]{Planck:2019nip}%
  \BibitemOpen
  \bibfield  {author} {\bibinfo {author} {\bibfnamefont {N.}~\bibnamefont
  {Aghanim}} \emph {et~al.} (\bibinfo {collaboration} {Planck}),\ }\href
  {\doibase 10.1051/0004-6361/201936386} {\bibfield  {journal} {\bibinfo
  {journal} {Astron. Astrophys.}\ }\textbf {\bibinfo {volume} {641}},\ \bibinfo
  {pages} {A5} (\bibinfo {year} {2020}{\natexlab{b}})},\ \Eprint
  {http://arxiv.org/abs/1907.12875} {arXiv:1907.12875 [astro-ph.CO]}
  \BibitemShut {NoStop}%
\bibitem [{\citenamefont {Aghanim}\ \emph
  {et~al.}(2020{\natexlab{c}})\citenamefont {Aghanim} \emph
  {et~al.}}]{Planck:2018lbu}%
  \BibitemOpen
  \bibfield  {author} {\bibinfo {author} {\bibfnamefont {N.}~\bibnamefont
  {Aghanim}} \emph {et~al.} (\bibinfo {collaboration} {Planck}),\ }\href
  {\doibase 10.1051/0004-6361/201833886} {\bibfield  {journal} {\bibinfo
  {journal} {Astron. Astrophys.}\ }\textbf {\bibinfo {volume} {641}},\ \bibinfo
  {pages} {A8} (\bibinfo {year} {2020}{\natexlab{c}})},\ \Eprint
  {http://arxiv.org/abs/1807.06210} {arXiv:1807.06210 [astro-ph.CO]}
  \BibitemShut {NoStop}%
\bibitem [{\citenamefont {Beutler}\ \emph {et~al.}(2011)\citenamefont
  {Beutler}, \citenamefont {Blake}, \citenamefont {Colless}, \citenamefont
  {Jones}, \citenamefont {Staveley-Smith}, \citenamefont {Campbell},
  \citenamefont {Parker}, \citenamefont {Saunders},\ and\ \citenamefont
  {Watson}}]{beutler20116df}%
  \BibitemOpen
  \bibfield  {author} {\bibinfo {author} {\bibfnamefont {F.}~\bibnamefont
  {Beutler}}, \bibinfo {author} {\bibfnamefont {C.}~\bibnamefont {Blake}},
  \bibinfo {author} {\bibfnamefont {M.}~\bibnamefont {Colless}}, \bibinfo
  {author} {\bibfnamefont {D.~H.}\ \bibnamefont {Jones}}, \bibinfo {author}
  {\bibfnamefont {L.}~\bibnamefont {Staveley-Smith}}, \bibinfo {author}
  {\bibfnamefont {L.}~\bibnamefont {Campbell}}, \bibinfo {author}
  {\bibfnamefont {Q.}~\bibnamefont {Parker}}, \bibinfo {author} {\bibfnamefont
  {W.}~\bibnamefont {Saunders}}, \ and\ \bibinfo {author} {\bibfnamefont
  {F.}~\bibnamefont {Watson}},\ }\href@noop {} {\bibfield  {journal} {\bibinfo
  {journal} {Monthly Notices of the Royal Astronomical Society}\ }\textbf
  {\bibinfo {volume} {416}},\ \bibinfo {pages} {3017} (\bibinfo {year}
  {2011})}\BibitemShut {NoStop}%
\bibitem [{\citenamefont {Ross}\ \emph {et~al.}(2015)\citenamefont {Ross},
  \citenamefont {Samushia}, \citenamefont {Howlett}, \citenamefont {Percival},
  \citenamefont {Burden},\ and\ \citenamefont {Manera}}]{Ross:2014qpa}%
  \BibitemOpen
  \bibfield  {author} {\bibinfo {author} {\bibfnamefont {A.~J.}\ \bibnamefont
  {Ross}}, \bibinfo {author} {\bibfnamefont {L.}~\bibnamefont {Samushia}},
  \bibinfo {author} {\bibfnamefont {C.}~\bibnamefont {Howlett}}, \bibinfo
  {author} {\bibfnamefont {W.~J.}\ \bibnamefont {Percival}}, \bibinfo {author}
  {\bibfnamefont {A.}~\bibnamefont {Burden}}, \ and\ \bibinfo {author}
  {\bibfnamefont {M.}~\bibnamefont {Manera}},\ }\href {\doibase
  10.1093/mnras/stv154} {\bibfield  {journal} {\bibinfo  {journal} {Mon. Not.
  Roy. Astron. Soc.}\ }\textbf {\bibinfo {volume} {449}},\ \bibinfo {pages}
  {835} (\bibinfo {year} {2015})},\ \Eprint {http://arxiv.org/abs/1409.3242}
  {arXiv:1409.3242 [astro-ph.CO]} \BibitemShut {NoStop}%
\bibitem [{\citenamefont {Riess}\ \emph {et~al.}(2022)\citenamefont {Riess}
  \emph {et~al.}}]{Riess:2021jrx}%
  \BibitemOpen
  \bibfield  {author} {\bibinfo {author} {\bibfnamefont {A.~G.}\ \bibnamefont
  {Riess}} \emph {et~al.},\ }\href {\doibase 10.3847/2041-8213/ac5c5b}
  {\bibfield  {journal} {\bibinfo  {journal} {Astrophys. J. Lett.}\ }\textbf
  {\bibinfo {volume} {934}},\ \bibinfo {pages} {L7} (\bibinfo {year} {2022})},\
  \Eprint {http://arxiv.org/abs/2112.04510} {arXiv:2112.04510 [astro-ph.CO]}
  \BibitemShut {NoStop}%
\bibitem [{\citenamefont {Di~Valentino}\ \emph
  {et~al.}(2021{\natexlab{a}})\citenamefont {Di~Valentino}, \citenamefont
  {Mena}, \citenamefont {Pan}, \citenamefont {Visinelli}, \citenamefont {Yang},
  \citenamefont {Melchiorri}, \citenamefont {Mota}, \citenamefont {Riess},\
  and\ \citenamefont {Silk}}]{DiValentino:2021izs}%
  \BibitemOpen
  \bibfield  {author} {\bibinfo {author} {\bibfnamefont {E.}~\bibnamefont
  {Di~Valentino}}, \bibinfo {author} {\bibfnamefont {O.}~\bibnamefont {Mena}},
  \bibinfo {author} {\bibfnamefont {S.}~\bibnamefont {Pan}}, \bibinfo {author}
  {\bibfnamefont {L.}~\bibnamefont {Visinelli}}, \bibinfo {author}
  {\bibfnamefont {W.}~\bibnamefont {Yang}}, \bibinfo {author} {\bibfnamefont
  {A.}~\bibnamefont {Melchiorri}}, \bibinfo {author} {\bibfnamefont {D.~F.}\
  \bibnamefont {Mota}}, \bibinfo {author} {\bibfnamefont {A.~G.}\ \bibnamefont
  {Riess}}, \ and\ \bibinfo {author} {\bibfnamefont {J.}~\bibnamefont {Silk}},\
  }\href {\doibase 10.1088/1361-6382/ac086d} {\bibfield  {journal} {\bibinfo
  {journal} {Class. Quant. Grav.}\ }\textbf {\bibinfo {volume} {38}},\ \bibinfo
  {pages} {153001} (\bibinfo {year} {2021}{\natexlab{a}})},\ \Eprint
  {http://arxiv.org/abs/2103.01183} {arXiv:2103.01183 [astro-ph.CO]}
  \BibitemShut {NoStop}%
\bibitem [{\citenamefont {Abdalla}\ \emph {et~al.}(2022)\citenamefont {Abdalla}
  \emph {et~al.}}]{Abdalla:2022yfr}%
  \BibitemOpen
  \bibfield  {author} {\bibinfo {author} {\bibfnamefont {E.}~\bibnamefont
  {Abdalla}} \emph {et~al.},\ }\href {\doibase 10.1016/j.jheap.2022.04.002}
  {\bibfield  {journal} {\bibinfo  {journal} {JHEAp}\ }\textbf {\bibinfo
  {volume} {34}},\ \bibinfo {pages} {49} (\bibinfo {year} {2022})},\ \Eprint
  {http://arxiv.org/abs/2203.06142} {arXiv:2203.06142 [astro-ph.CO]}
  \BibitemShut {NoStop}%
\bibitem [{\citenamefont {Di~Valentino}\ \emph
  {et~al.}(2021{\natexlab{b}})\citenamefont {Di~Valentino} \emph
  {et~al.}}]{DiValentino:2020vvd}%
  \BibitemOpen
  \bibfield  {author} {\bibinfo {author} {\bibfnamefont {E.}~\bibnamefont
  {Di~Valentino}} \emph {et~al.},\ }\href {\doibase
  10.1016/j.astropartphys.2021.102604} {\bibfield  {journal} {\bibinfo
  {journal} {Astropart. Phys.}\ }\textbf {\bibinfo {volume} {131}},\ \bibinfo
  {pages} {102604} (\bibinfo {year} {2021}{\natexlab{b}})},\ \Eprint
  {http://arxiv.org/abs/2008.11285} {arXiv:2008.11285 [astro-ph.CO]}
  \BibitemShut {NoStop}%
\end{thebibliography}%

\end{document}